# Three-dimensional metamaterials with an ultra-high effective refractive index over broad bandwidth

Jonghwa Shin[*], Jung-Tsung Shen[†], and Shanhui Fan[‡]

*Ginzton Lab and Department of Electrical Engineering, Stanford University, Stanford, California 94305*

The authors introduce a general mechanism, based on electrostatic and magnetostatic considerations, for designing three-dimensional isotopic metamaterials that possess an enhanced refractive index over an extremely large frequency range. The mechanism allows nearly independent control of effective electric permittivity and magnetic permeability without the use of resonant elements.

One of the motivations for developing metamaterials is to achieve relative electric permittivity $\varepsilon_r$ and magnetic permeability $\mu_r$, in the ranges not readily accessible using naturally occurring materials [1]. In particular, creating an arbitrarily high refractive index ($n = \sqrt{\varepsilon_r \mu_r}$) is of interest for imaging and lithography, where the resolution scales inversely with the refractive index [2]. Moreover, increasing the refractive index over a large frequency range results in broadband slow light, which can be used to enhance the storage capacity of delay lines [3] as well as spectral sensitivity in interferometers [4].

In this Letter, we design three-dimensional metamaterials with an index of refraction that is arbitrarily high and nondispersive, over a broad frequency range extending down to near-zero frequency. In contrast to our work, previous approaches to enhancing the refractive index utilized either electronic resonances in atoms [5] or electromagnetic resonances such as split-ring resonators [6,7]. These schemes are inherently narrowband and work only in the vicinity of a resonant frequency. Related to our work, it was discovered that the use of an array of subwavelength capacitors could lead to the frequency-independent and broadband enhancement of the relative electric permittivity $\varepsilon_r$ [8–10]. However, all of the previously studied capacitive metamaterial structures exhibited strong diamagnetic behavior that suppressed the relative magnetic permeability $\mu_r$ in the low frequency regime [10–12]. Hence, in all previous systems, the capability of increasing the refractive index *n* was very limited.

Here we propose a general mechanism to greatly enhance the $\varepsilon_r$ of a metamaterial without suppressing $\mu_r$. The key idea is to generate a large electric dipole response, while simultaneously preventing the formation of large-area current loops. Below, we introduce the details of our mechanism using three crystal designs with progressive structural changes. All these structures consist of a cubic array of isolated perfect electric conductor (PEC) objects and have either an octahedral ($O_h$) or a pyritohedral ($T_h$) symmetry point group. Hence, their electromagnetic properties are isotropic [13,14], with $\varepsilon_r$ and $\mu_r$ being scalars.

To begin with, we consider a simple cubic array structure of metal cubes [Fig. 1(a)]. For a structure with $b=19a/20$, where *a* and *b* are the linear sizes of the unit cell and the metal cube, respectively, the transmission spectrum through a slab with a thickness of $10a$ is shown in the bottom panel of Fig. 1(a). These spectra were calculated using finite difference time domain (FDTD) schemes [15]. Accurate values of effective $\varepsilon_r$ and $\mu_r$ can be extracted from such transmission spectra: the frequency spacing between the transmission peaks reveals the index of refraction, while the impedance can be found from the minimum value of the transmission coefficient. (Alternatively, we have also applied the s-parameter extraction method [16] to the transmission and reflection coefficients and obtained the same $\varepsilon_r$ and $\mu_r$.). The permittivity and permeability are found to be $\varepsilon_r$=20.0 and $\mu_r$=0.098, yielding n=1.4. While this structure provides $\varepsilon_r$ enhancement, it exhibits strong diamagnetism that suppresses $\mu_r$. As a result, the structure demonstrates only a relatively modest increase in the refractive index, in agreement with Ref. 12.

Since the structure is isotropic, to develop a physical intuition about its behavior it is sufficient to study its response to externally applied fields in the *z*-direction. For this structure, the opposing faces of the nearest-neighbor cubes form parallel-plate capacitors. Applying an electric field in the *z*-direction creates a strong field within such capacitors with *z*-normal faces, as well as a large accumulation of surface charges [top panel of Fig. 2(a)], resulting in a substantial dipole moment and large electric permittivity. More specifically, the relative permittivity can be calculated with an area average, on the *x*–*y* plane, of the electric field in the gap region, divided by a line average of electric field along the *z*-direction [6]. Therefore, for large cubes almost filling the entire unit cell (i.e., $b \approx a$), the relative permittivity of the metal cube structure is approximately given by *a*/(*a-b*).

In accordance with Lenz's law, a quasi-static (but still time-varying) magnetic field induces electric current [middle panel of Fig. 2(a)] on the surface of PEC, such that the magnetic field vanishes within [bottom panel of Fig. 2(a)]. The magnetic moments generated by these electric currents therefore align in the opposite direction to the applied magnetic field, making the structure diamagnetic. When the cubes are large, i.e., $b \approx a$, the small gap that is normal to the field can be ignored in magnetic response calculations and the structure can be approximated by a two-dimensional lattice of pillars of infinite length. For such pillar structures, the permeability in the direction of the pillar axis is $\mu_r = 1 - A_{\text{pillar}}/A_{\text{unit}}$, where $A_{\text{pillar}}$ and $A_{\text{unit}}$ are the cross-sectional areas of the pillar and the unit cell, respectively. Hence, for the metal cube structure in the large cube limit, the magnetic permeability becomes $1 - b^2/a^2 \approx 2(a-b)/a$, which agrees with the numerical analysis. In this structure, the enhancement of dielectric permittivity is also associated with a strong diamagnetic response that suppresses $\mu_r$. Such a strong diamagnetic response, in fact, holds true for all proposed metamaterial structures with a strong capacitative response [10–12] and limits the capability of index enhancement.

To reduce the diamagnetic response of the structure, we note from the discussions above that $\varepsilon_r$ and $\mu_r$ are controlled by different aspects of the structure. The permittivity is determined by the charges on the surfaces normal to the electric field, while the permeability is determined by the current on the surfaces parallel to the magnetic field. The strength of magnetic dipole moment is defined as the product of the current and the area enclosed by the current loop. Thus, to create a structure with an enhanced index, one should seek to reduce the area of the current loops in the system while maintaining a strong capacitative response.

For *z*-directional fields, one way to decrease the current loop area, without affecting the electric response, is to replace the cube with two *z*-normal metal plates connected by a metal wire [Fig. 2(b)]. (The connecting wire is very important as it maintains the two plates at an equal electric potential. Without it, $\varepsilon_r$ reduces almost to unity for thin plates.) Comparing the top panels of Figs. 2(a) and (b), it is apparent that these two structures have an almost identical electric response since the surfaces normal to the electric field is identical. On the other hand, the magnetic response would be strongly affected by such a structural change since most of the current loops are now much smaller [middle and bottom panels of Fig. 2(b)].

As a direct numerical test of the concept outlined above, we simulate an isotropic structure in which the unit cell comprises six metal plates, two in each orthogonal direction [Fig. 1(b)]. Three orthogonal metal wires intersecting at the center of the unit cell connect all six plates. We designed the metal element, with the same outer dimension as the cube of the metal cube lattice structure, i.e., $b=19a/20$. For this structure, $\varepsilon_r$ and $\mu_r$ are found to be 19.6 and 0.568, respectively, from the transmission spectrum [bottom panel of Fig. 1(b)], and yield a refractive index $n$ of 3.34. Thus, while the permittivity of the structure is nearly identical to that of the cube structure, the permeability is increased more than fivefold.

Further suppression of the diamagnetic response can be accomplished by introducing air slits into the plates [Fig. 1(c)], which causes additional reduction of the area of the current loops. The operation of this structure can be visualized again by focusing on the case in which external fields are aligned in the $z$-direction of a simplified structure shown in Fig. 2(c), with only two metal plates. As a result of the reduction in the area of the current loops in the plates [middle panel of Fig. 2(c)], the magnetic field can now penetrate much deeper into the structure, which signifies the suppression of the diamagnetic response. [bottom panel of Fig. 2(c)]. On the other hand, the permittivity does not change significantly as long as the slits are narrow enough so that the electric fields are blocked from passing through. The charge distribution is now non-uniform on the metal plates, with the high concentration of charges near the edges of the slits compensating for those charges that were previously on the surface having been removed, and the total amount of surface charge in the top panel of Fig. 2(c) remains relatively unchanged from Fig. 2(b). (This is, in fact, a fringe-field effect. It is well-known that to describe the capacitance of a finite-size parallel plate capacitor, one needs to use an effective area of the plate that is larger than its physical size. [17,18].) The intuition above is confirmed numerically for the isotropic structure shown in Fig. 1(c), as FDTD simulations show $\varepsilon_r$=18.3 and $\mu_r$=0.966, resulting in a refractive index of 4.2. Thus, the structure in Fig. 1(c) has approximately the same electric response as the cube structure in Fig. 1(a), but almost no magnetic response.

Starting from a structure similar to Fig. 1(c), one can further enhance the index by decreasing the distance between the metal plates, and simultaneously reducing the thickness of plates, and the width and spacing of slits, while increasing the number of slits in the plates. For demonstration, we chose the thickness of the plates, and the width and spacing between the slits, as well as the distance between the plates all to be equal to (*a-b*), and compared different structures as we changed *a-b*. The semi-analytic derivation [19] predicts that the effective $\varepsilon_r$ scale linearly with *a*/(*a-b*), while the effective $\mu_r$ approaches more closely to unity. Numerical simulation results, as shown in Fig. 3, agree very well with such predictions. This simulation demonstrates the capability of creating an ultra-high index metamaterial.

Finally, our results suggest a general mechanism to control the permittivity and permeability of a metamaterial independently. It is also noteworthy that the enhancement of permittivity and the suppression of diamagnetic behaviors, as described here, only results from the *shape* of the metal objects and its relative size to the unit cell, and are independent of the absolute size of the unit cell or the object, as long as these objects and the periodicity of the crystal stay within the deep sub-wavelength limit. Consequently, within the deep sub-wavelength limit, the effective permittivity and permeability do not depend on the frequency of the electromagnetic waves. The properties of these structures can, therefore, be designed to be non-dispersive over a very broad frequency range, extending from a near-zero frequency to the mid-infrared range, as long as the metal can be well approximated as near-perfect electric conductors.

This work is supported in part by ARO (Grant No. DAAD-19-03-1-0227) and AFOSR (Grant No. FA9550-04-1-0437), and the DARPA Slow Light Program (Grant No. FA9550-04-0414). The simulations were carried out in the Pittsburgh Supercomputing Center through the support of the NSF LRAC.

# References

*Electronic mail: jshin@jshin.info

†Electronic mail: jushen@gmail.com

‡Electronic mail: shanhui@stanford.edu

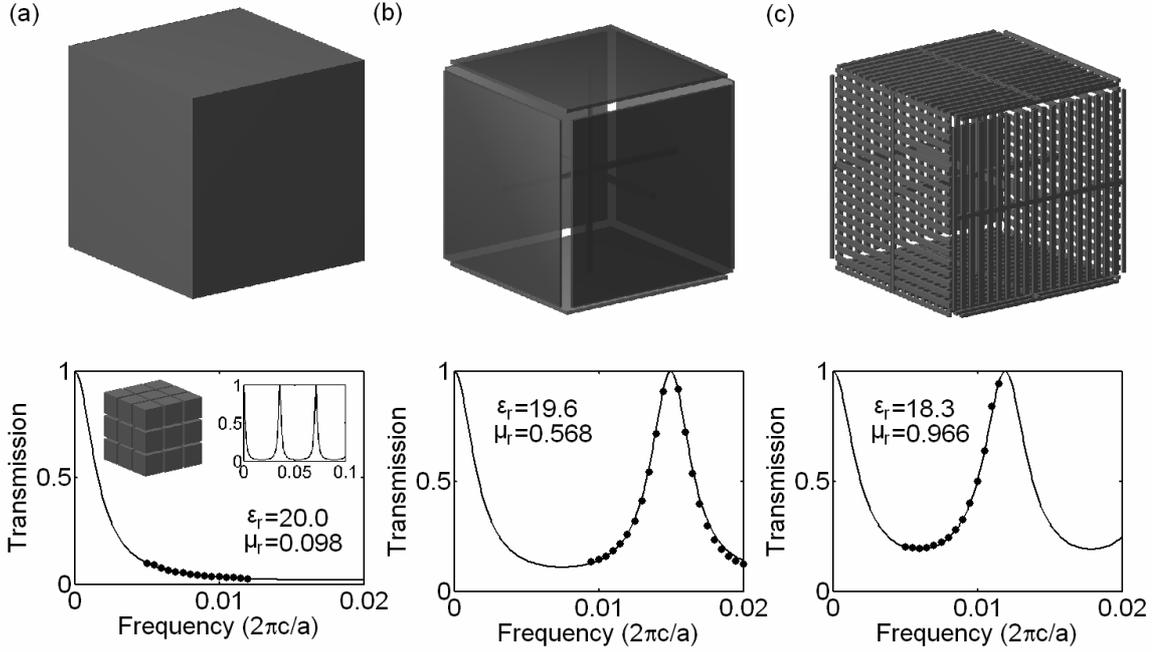

FIG 1. Electromagnetic responses of metamaterials consisting of a cubic lattice of metallic objects. In all three structures, the outer dimension of the metallic objects is $19a/20$, where $a$ is the lattice constant. In (a)–(c), the top panels show the object and the bottom panels are the corresponding transmission spectra through a slab of such metamaterial with a thickness of $10a$. The dots are from FDTD simulation and the lines are the response of a corresponding uniform medium with an effective $\varepsilon$ and $\mu$ indicated. (a) Metallic cube. The left and right insets in the bottom panel show a structure with 3-by-3-by-3 unit cells, and a transmission spectrum over broader frequency range, respectively. (b) A cube with connected metal plates. (c) Similar to (b), except that each plate now has slits.

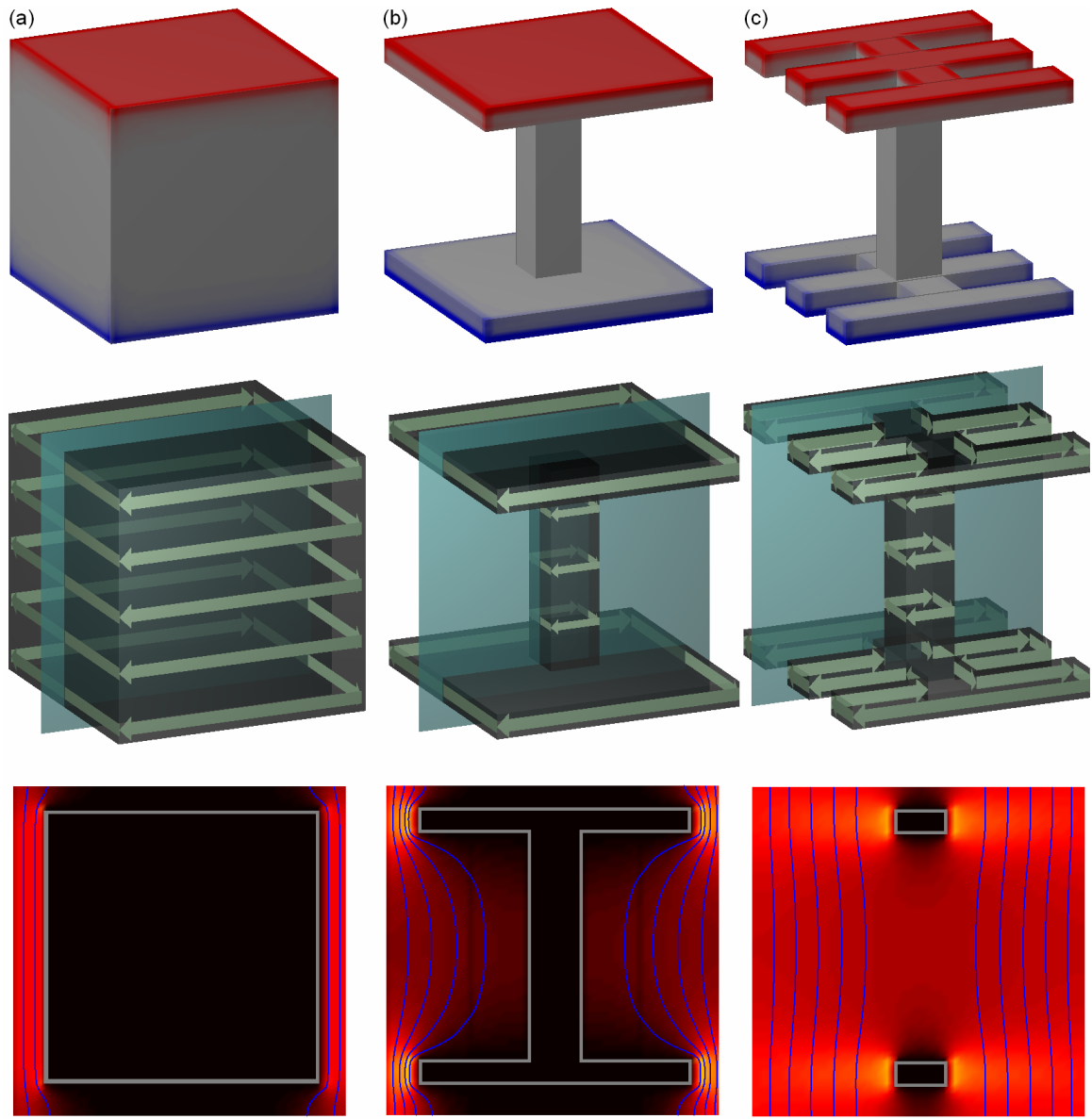

FIG 2. Electromagnetic response of metallic objects to externally applied electrical and magnetic fields along the z-direction. The top panel in (a)–(c) shows the object, the colors indicate the charge distribution when the electric field is applied. The middle panel in (a)–(c) shows the current distribution when the magnetic field is applied. The bottom panels show the distribution and the direction of the total magnetic field on a slice as indicated in the middle panels. Gray lines indicate the boundaries of the metallic region. (a) a metallic cube. (b) two metallic plates connected by a metal-wire. (c) similar to (b), except that each plate now has slits.

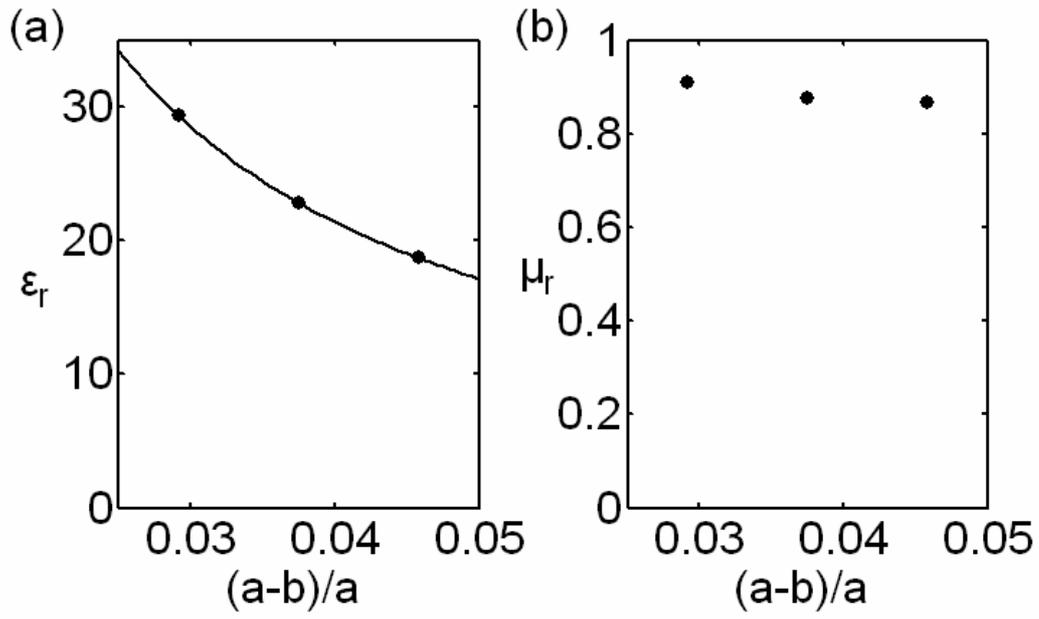

FIG 3. Numerical results of (a) relative permittivity and (b) relative permeability as a function of $a-b$ in structures similar to Fig. 1(c), where $b$ is the outer dimension of the metallic object, and $a$ is the periodicity. The solid line is a theoretical prediction.